


\font\twelverm=cmr10  scaled 1200   \font\twelvei=cmmi10  scaled 1200
\font\twelvesy=cmsy10 scaled 1200   \font\twelveex=cmex10 scaled 1200
\font\twelvebf=cmbx10 scaled 1200   \font\twelvesl=cmsl10 scaled 1200
\font\twelvett=cmtt10 scaled 1200   \font\twelveit=cmti10 scaled 1200
\font\twelvesc=cmcsc10 scaled 1200
\skewchar\twelvei='177   \skewchar\twelvesy='60


\def\twelvepoint{\normalbaselineskip=12.4pt plus 0.1pt minus 0.1pt
  \abovedisplayskip 12.4pt plus 3pt minus 9pt
  \belowdisplayskip 12.4pt plus 3pt minus 9pt
  \abovedisplayshortskip 0pt plus 3pt
  \belowdisplayshortskip 7.2pt plus 3pt minus 4pt
  \smallskipamount=3.6pt plus1.2pt minus1.2pt
  \medskipamount=7.2pt plus2.4pt minus2.4pt
  \bigskipamount=14.4pt plus4.8pt minus4.8pt
  \def\rm{\fam0\twelverm}          \def\it{\fam\itfam\twelveit}%
  \def\sl{\fam\slfam\twelvesl}     \def\bf{\fam\bffam\twelvebf}%
  \def\mit{\fam 1}                 \def\cal{\fam 2}%
  \def\sc{\twelvesc}               \def\tt{\twelvett}
  \def\sf{\twelvesf}
  \textfont0=\twelverm   \scriptfont0=\tenrm   \scriptscriptfont0=\sevenrm
  \textfont1=\twelvei    \scriptfont1=\teni    \scriptscriptfont1=\seveni
  \textfont2=\twelvesy   \scriptfont2=\tensy   \scriptscriptfont2=\sevensy
  \textfont3=\twelveex   \scriptfont3=\twelveex  \scriptscriptfont3=\twelveex
  \textfont\itfam=\twelveit
  \textfont\slfam=\twelvesl
  \textfont\bffam=\twelvebf \scriptfont\bffam=\tenbf
  \scriptscriptfont\bffam=\sevenbf
  \normalbaselines\rm}



\def\beginlinemode{\endmode
  \begingroup\parskip=0pt \obeylines\def\\{\par}\def\endmode{\par\endgroup}}
\def\beginparmode{\endmode
  \begingroup \def\endmode{\par\endgroup}}
\let\endmode=\par
{\obeylines\gdef\
{}}
\def\singlespace{\baselineskip=\normalbaselineskip}

\def\oneandahalfspace{\baselineskip=\normalbaselineskip
  \multiply\baselineskip by 3 \divide\baselineskip by 2}
\def\doublespace{\baselineskip=\normalbaselineskip \multiply\baselineskip by 2}

\newcount\firstpageno
\firstpageno=2
\footline={\ifnum\pageno<\firstpageno{\hfil}\else{\hfil\twelverm\folio\hfil}\fi}

\def\toppageno{\global\footline={\hfil}\global\headline
  ={\ifnum\pageno<\firstpageno{\hfil}\else{\hfil\twelverm\folio\hfil}\fi}}
\let\rawfootnote=\footnote              
\def\footnote#1#2{{\rm\singlespace\parindent=0pt\parskip=0pt
  \rawfootnote{#1}{#2\hfill\vrule height 0pt depth 6pt width 0pt}}}
\def\raggedcenter{\leftskip=4em plus 12em \rightskip=\leftskip
  \parindent=0pt \parfillskip=0pt \spaceskip=.3333em \xspaceskip=.5em
  \pretolerance=9999 \tolerance=9999
  \hyphenpenalty=9999 \exhyphenpenalty=9999 }
\def\dateline{\rightline{\ifcase\month\or
  January\or February\or March\or April\or May\or June\or
  July\or August\or September\or October\or November\or December\fi
  \space\number\year}}
\def\received{\vskip 3pt plus 0.2fill
 \centerline{\sl (Received\space\ifcase\month\or
  January\or February\or March\or April\or May\or June\or
  July\or August\or September\or October\or November\or December\fi
  \qquad, \number\year)}}


\hsize=6.5truein
\hoffset=0.0truein
\vsize=8.5truein
\voffset=0.25truein
\parskip=\medskipamount
\toppageno
\twelvepoint
\doublespace
\def\\{\cr}
\overfullrule=0pt 




\def\title#1{                   
   \null \vskip 3pt plus 0.3fill \beginlinemode
   \doublespace \raggedcenter {\bf #1} \vskip 3pt plus 0.1 fill}

\def\author                     
  {\vskip 3pt plus 0.1fill \beginlinemode \doublespace \raggedcenter}

\def\affil                      
  {\vskip 3pt \beginlinemode \doublespace \raggedcenter \it}

\def\abstract                   
  {\vskip 3pt plus 0.1fill \subhead {Abstract:}
   \beginparmode \narrower \oneandahalfspace }

\def\endtopmatter               
  {\vskip 3pt plus 0.1fill \endpage \body}

\def\body                       
  {\beginparmode}               

\def\head#1{                    
   \goodbreak \vskip 0.4truein  
  {\immediate\write16{#1} \raggedcenter {\sc #1} \par}
   \nobreak \vskip 3pt \nobreak}

\def\subhead#1{                 
  \vskip 0.25truein             
  {\raggedcenter {\it #1} \par} \nobreak \vskip 3pt \nobreak}

\def\beneathrel#1\under#2{\mathrel{\mathop{#2}\limits_{#1}}}

\def\refto#1{${\,}^{#1}$}       

\newdimen\refskip \refskip=0pt
\def\references         
  {\head{References}    
   \beginparmode \frenchspacing \parindent=0pt \leftskip=\refskip
   \parskip=0pt \everypar{\hangindent=20pt\hangafter=1}}

\gdef\refis#1{\item{#1.\ }}                     

\gdef\journal#1, #2, #3 {               
    {\it #1}, {\bf #2}, #3.}            




\def\endreferences{\body}

\def\figurecaptions             
  {\endpage \beginparmode \head{Figure Captions}
   \parskip=3pt \everypar{\hangindent=20pt\hangafter=1} }

\def\endpage                    
  {\vfill\eject}

\def\endpaper   {\endmode\vfill\supereject}
\def\endjnl     {\endpaper\end}


\def\ref#1{ref.{#1}}                    
\def\Ref#1{Ref.{#1}}                    
\def\[#1]{[\cite{#1}]}
\def\cite#1{{#1}}


\def\(#1){(\call{#1})}
\def\call#1{{#1}}
\def\frac#1#2{{#1 \over #2}}

\def\12{{1\over2}}

\def\sla{\raise.15ex\hbox{$/$}\kern-.57em}
\def\leaderfill{\leaders\hbox to 1em{\hss.\hss}\hfill}
\def\twiddle{\lower.9ex\rlap{$\kern-.1em\scriptstyle\sim$}}
\def\bigtwiddle{\lower1.ex\rlap{$\sim$}}
\def\gtwid{\mathrel{\raise.3ex\hbox{$>$\kern-.75em\lower1ex\hbox{$\sim$}}}}
\def\ltwid{\mathrel{\raise.3ex\hbox{$<$\kern-.75em\lower1ex\hbox{$\sim$}}}}
\def\square{\kern1pt\vbox{\hrule height 1.2pt\hbox{\vrule width 1.2pt\hskip 3pt
   \vbox{\vskip 6pt}\hskip 3pt\vrule width 0.6pt}\hrule height 0.6pt}\kern1pt}
\def\tdot#1{\mathord{\mathop{#1}\limits^{\kern2pt\ldots}}}

\def\pmb#1{\setbox0=\hbox{#1}%
  \kern-.025em\copy0\kern-\wd0
  \kern  .05em\copy0\kern-\wd0
  \kern-.025em\raise.0433em\box0 }

\catcode`@=11
\newcount\r@fcount \r@fcount=0
\newcount\r@fcurr
\immediate\newwrite\reffile
\newif\ifr@ffile\r@ffilefalse
\def\w@rnwrite#1{\ifr@ffile\immediate\write\reffile{#1}\fi\message{#1}}

\def\writer@f#1>>{}
\def\referencefile{
  \r@ffiletrue\immediate\openout\reffile=\jobname.ref%
  \def\writer@f##1>>{\ifr@ffile\immediate\write\reffile%
    {\noexpand\refis{##1} = \csname r@fnum##1\endcsname = %
     \expandafter\expandafter\expandafter\strip@t\expandafter%
     \meaning\csname r@ftext\csname r@fnum##1\endcsname\endcsname}\fi}%
  \def\strip@t##1>>{}}

\def\citeall#1{\xdef#1##1{#1{\noexpand\cite{##1}}}}
\def\cite#1{\each@rg\citer@nge{#1}}     

\def\each@rg#1#2{{\let\thecsname=#1\expandafter\first@rg#2,\end,}}
\def\first@rg#1,{\thecsname{#1}\apply@rg}       
\def\apply@rg#1,{\ifx\end#1\let\next=\relax
\else,\thecsname{#1}\let\next=\apply@rg\fi\next}

\def\citer@nge#1{\citedor@nge#1-\end-}  
\def\citer@ngeat#1\end-{#1}
\def\citedor@nge#1-#2-{\ifx\end#2\r@featspace#1 
  \else\citel@@p{#1}{#2}\citer@ngeat\fi}        
\def\citel@@p#1#2{\ifnum#1>#2{\errmessage{Reference range #1-#2\space is bad.}
    \errhelp{If you cite a series of references by the notation M-N, then M and
    N must be integers, and N must be greater than or equal to M.}}\else%
 {\count0=#1\count1=#2\advance\count1
by1\relax\expandafter\r@fcite\the\count0,%
  \loop\advance\count0 by1\relax
    \ifnum\count0<\count1,\expandafter\r@fcite\the\count0,%
  \repeat}\fi}

\def\r@featspace#1#2 {\r@fcite#1#2,}    
\def\r@fcite#1,{\ifuncit@d{#1}          
    \expandafter\gdef\csname r@ftext\number\r@fcount\endcsname%
    {\message{Reference #1 to be supplied.}\writer@f#1>>#1 to be supplied.\par
     }\fi%
  \csname r@fnum#1\endcsname}

\def\ifuncit@d#1{\expandafter\ifx\csname r@fnum#1\endcsname\relax%
\global\advance\r@fcount by1%
\expandafter\xdef\csname r@fnum#1\endcsname{\number\r@fcount}}

\let\r@fis=\refis                       
\def\refis#1#2#3\par{\ifuncit@d{#1}
    \w@rnwrite{Reference #1=\number\r@fcount\space is not cited up to now.}\fi%
  \expandafter\gdef\csname r@ftext\csname r@fnum#1\endcsname\endcsname%
  {\writer@f#1>>#2#3\par}}

\def\r@ferr{\endreferences\errmessage{I was expecting to see
\noexpand\endreferences before now;  I have inserted it here.}}
\let\r@ferences=\references
\def\references{\r@ferences\def\endmode{\r@ferr\par\endgroup}}

\let\endr@ferences=\endreferences
\def\endreferences{\r@fcurr=0
  {\loop\ifnum\r@fcurr<\r@fcount
    \advance\r@fcurr by 1\relax\expandafter\r@fis\expandafter{\number\r@fcurr}%
    \csname r@ftext\number\r@fcurr\endcsname%
  \repeat}\gdef\r@ferr{}\endr@ferences}


\let\r@fend=\endpaper\gdef\endpaper{\ifr@ffile
\immediate\write16{Cross References written on []\jobname.REF.}\fi\r@fend}

\catcode`@=12

\citeall\refto          
\citeall\ref            %
\citeall\Ref            %

\doublespace
\vglue 0. truein
\title
{
Electroweak String Configurations with Baryon Number
}
\smallskip
\author
{Tanmay Vachaspati}
\affil
{
Tufts Institute of Cosmology, Department of Physics and Astronomy,
Tufts University, Medford, MA 02155.
}
\smallskip
\author
{George B. Field}
\affil
{
Harvard-Smithsonian Center for Astrophysics,
60 Garden Street, Cambridge, MA 02138.
}

\abstract
\doublespace

In the context of electroweak strings, the baryon number anomaly equation
may be reinterpreted as a conservation law for baryon number minus helicity.
Since the helicity is a sum of the link and twist numbers, linked or twisted
loops of electroweak string carry baryon number. We evaluate the
change in the baryon number obtained by delinking loops of electroweak
$Z-$string and show that twisted electroweak string segments may be regarded
as extended sphalerons. We also suggest an alternative scenario for
electroweak baryogenesis.

\endtopmatter

Over the past few years, there has been renewed interest in the study
of classical solutions in the standard model of the electroweak
interactions.
Vortex solutions\refto{yn, nm, tv1, tvmb} are of particular interest and
there is indication\refto{tv, mhmj, mbtvmb}
that they may be the building blocks for other solutions, such as the
sphaleron\refto{nm, fknm}.
The sphaleron, in turn, is of crucial interest to the study of baryon
number violating processes and to the possibility of baryogenesis
during the cosmological electroweak phase transition\refto{nt, review}.
The point of this paper is to show that electroweak string configurations
can carry baryon number and play a (sphaleron-like) role in baryon number
changing processes.

The starting point for our analysis is the anomaly equation:
$$
\partial_\mu j^\mu _{B} = {{N_F} \over {32\pi^2}}
 [ -g^2 W^a _{\mu \nu} {\tilde W}^{a \mu \nu} +
                         {g'}{}^2 Y_{\mu \nu} {\tilde Y}^{\mu \nu} ].
\eqno (1)
$$
in the usual notation (see Ref. \cite{nt} for example).
The right-hand side of (1) is a total divergence and so the equation
can be integrated. If we assume that the baryonic flux through the
surface of the three volume of interest vanishes, the result relates
the change in the baryon number within the volume, $Q_B$, to the Chern-Simons
numbers of the fields:
$$
\Delta Q_B = N_F \Delta ( N_{CS} - n_{CS} )
\eqno (2)
$$
where, the $SU(2)_L$ Chern-Simons number is,
$$
N_{CS} \equiv {{g^2 } \over {32\pi^2}}  \int  d^3 x \epsilon_{ijk}
 \biggl [  W^{a ij} W^{a k} - {g \over 3}
                  \epsilon_{abc} W^{ai} W^{bj} W^{ck}   \biggr ]
\eqno (3)
$$
and, the $U(1)_Y$ Chern-Simons number is,
$$
n_{CS} \equiv {{{g'}{}^2 } \over {32\pi^2}}  \int  d^3 x \epsilon_{ijk}
 \biggl [  Y^{ij} Y^{k} \biggr ] \ .
\eqno (4)
$$
The $\Delta$ in (2) means that the difference is to be taken between
initial and final configurations; $i,j,k$ are spatial indices and
$a,b,c$ are group indices.

We will be interested in initial and final field configurations in which
$W^1 _\mu = 0 = W^2 _\mu$. With this simplification,
and the transformation,
$$
Z_j = cos\theta_w W^3_j - sin\theta_w Y_j \ , \ \ \
A_j = sin\theta_w W^3_j + cos\theta_w Y_j \ ,
\eqno (5)
$$
eqn. (2) gives,
$$
\Delta Q_B =\Delta \biggl [ N_F {{\alpha ^2 } \over {32\pi^2}} \int d^3 x
\biggl \{ cos(2\theta_w ) {\vec B}_Z \cdot {\vec Z} +
      {1 \over 2} sin(2\theta_w ) ( {\vec B}_Z \cdot {\vec A} +
                                     {\vec B}_A \cdot {\vec Z} )
\biggr \} \biggr ]
\eqno (6)
$$
where, $tan\theta_w = g'/g$, $\alpha = \sqrt{g^2 + {g'}{}^2}$, $\vec B$
denotes the magnetic field and the subscripts denote the gauge field
for which the magnetic field is to be evaluated.

The terms on the right-hand side have a simple interpretation
in terms of helicity\refto{mbgf}.
The helicity associated with the $Z$ field:
$$
H_Z = \int d^3 x {\vec B}_Z \cdot {\vec Z} \ .
\eqno (7)
$$
If we think in terms of flux tubes of $Z$ magnetic field, $H_Z$ measures
the sum of the link and twist number of these tubes\refto{comment10}:
$$
H_Z = L_Z + T_Z \ .
\eqno (8)
$$
For a pair of untwisted $Z$ flux tubes\refto{comment2}
that are linked once as shown in Fig. 1a,
the helicity is:
$$
H_Z =  2 \Phi_Z ^2
\eqno (9)
$$
where, $\Phi_Z$ is the magnetic flux in each of the two
tubes. Note that the helicity is positive for the strings shown in Fig. 1a.
If we reversed the direction of the flux in one of the loops, the magnitude
of $H_Z$ would be the same but the sign would change. For the $Z-$string,
we also know that
$$
\Phi_Z = {{4\pi} \over \alpha}
\eqno (10)
$$
and so eqn. (6) (ignoring the $\Delta$ sign for now) yields:
$$
CS (in) = N_F cos(2\theta_w ) \ .
\eqno (11)
$$
where $CS$ denotes the Chern-Simons number of the configuration.

Next consider the operation of delinking the loops shown in Fig. 1.
The first step is to let the loops self-intersect and intercommute. This
process preserves helicity\refto{tvintercom, comment5} as the linking of
the loops in Fig. 1a changes to a twist of the loop in Fig. 1b. The
twisted loop in Fig. 1b can be broken since the $Z-$string is not
topological. The result is a $Z-$string segment that is twisted by
$2\pi$ and has a monopole ($m$) at one end and an antimonopole ($\bar m$) at
the other (Fig. 1c). The field configurations of $m$ and $\bar m$ have
been written by Nambu\refto{yn, norm} :
$$
\Phi_m = \pmatrix{ cos(\theta_m /2) \cr sin(\theta_m /2) e^{i\phi }} \ ,
\ \ \
\Phi_{\bar m} = \pmatrix{ sin(\theta_{\bar m} /2) \cr
                          cos(\theta_{\bar m} /2) e^{i\phi} } \ .
\eqno (12)
$$
where, $\theta_m$ and $\phi$ are spherical coordinates centered on $m$ (and
similarly for $\bar m$).  In the $\theta_w = 0$ case, the gauge fields are
given by
$$
A_\mu = - i {2 \over g} \partial_\mu U \ U^{-1}
\eqno (13)
$$
where $U$ is a $2 \times 2$ matrix defined by $\Phi = U (1, 0)^T$.
(The case of non-zero $\theta_w$ requires a
more elaborate expression for the gauge field\refto{yn} and is treated
later.) The important thing to note is
that $-\Phi_m$ and $-\Phi_{\bar m}$ are also
valid Higgs field configurations for $m$ and $\bar m$ and
the gauge fields are unaffected by the overall minus sign. The next
step in the delinking process is shown in Fig. 1d, where the $Z-$string
segment of Fig. 1c is broken in the middle with $m$ and $\bar m$ in
the configurations $-\Phi_m$ and $-\Phi_{\bar m}$ respectively. Now
we have two $Z-$string segments, each one twisted by $\pi$ (Fig. 1d). The
next step is to perform rotations ($\phi \rightarrow \phi \pm \pi$)
of the newly created poles so that
the twists are undone (Fig. 1e). The Higgs configurations obtained this
way are called $-\Phi_{m} (- \pi )$ and $-\Phi_{\bar m} (\pi )$ in
Fig. 1e. Now we can write down the
Higgs configuration for one of the segments in Fig. 1e as:
$$
\Phi_{m \bar m} = \pmatrix{ cos\left (
                  {{\theta_m + \theta_{\bar m}} \over 2} \right ) \cr
      sin\left ( {{\theta_m + \theta_{\bar m}} \over 2} \right )
        e^{i\phi} }
\eqno (14)
$$
where, $\theta_m$ and $\theta_{\bar m}$ are measured from the $z-$axis
defined by the line from $\bar m$ to $m$.
This configuration has the right properties since
it reduces to $\Phi_m$ when $\theta_{\bar m} \rightarrow 0$ and it
reduces to $-\Phi_{\bar m} (\pi )$ when $\theta_m \rightarrow \pi$. It
also exhibits an untwisted $Z-$string segment between the poles. Now consider
what happens when the segment shrinks to zero size. Then
$\theta_m \rightarrow \theta_{\bar m} = \theta$, and we are left with
$$
\Phi_{m \bar m} = \pmatrix{ cos\theta \cr sin\theta e^{i\phi} } \ .
\eqno (15)
$$
But this, with the gauge fields given by (13), is precisely the Higgs
field configuration for
a sphaleron ($\Phi_s$) in the $\theta_w \rightarrow 0$ limit\refto{ed}.
Therefore the two segments of Fig. 1e shrink down to two sphalerons
(Fig. 1f) - each carrying $CS = N_F /2$ ($\theta_w = 0$) -
which can then decay into different vacuua with
$CS = 0, ~ N_F$ or $2N_F$. If we consider decay into the
$CS = 0$ vacuum, we must conclude that the linked loops of Fig. 1a
(with unit linkage) carry baryon number
$$
Q_B = N_F cos(2\theta_w ) \ .
\eqno (16)
$$
Note that we have explicitly shown
the equivalence of twisted electroweak strings with the sphaleron only in
the $\theta_w =0$ case; however, the result in (16) is true for any
$\theta_w$ because $CS(in)$ is given by (11) and the loops can always
decay into the vacuum with zero Chern-Simons number.

For general $\theta_w$, one can construct $Z-$string segments
with arbitrary twist and, equivalently, with arbitrary Chern-Simons
number. The secret to this construction is the realization that the
configurations $e^{i\gamma } \Phi_m$ and $e^{i\gamma } \Phi_{\bar m}$
also describe a monopole and an antimonopole for any constant
$\gamma$ since these are simply global gauge transformations of
$\Phi_m$ and $\Phi_{\bar m}$.
(For $\theta_w = 0$, we had chosen $\gamma = \pi$ to arrive
at (14)). Now consider the Higgs field configuration
$$
\Phi_{m \bar m} (\gamma ) =
\pmatrix{
            sin(\theta_m /2) sin(\theta_{\bar m} /2) e^{i\gamma}
          + cos(\theta_m /2) cos(\theta_{\bar m} /2) \cr
   sin(\theta_m /2) cos(\theta_{\bar m} /2) e^{i\phi }
 - cos(\theta_m /2) sin(\theta_{\bar m} /2) e^{i(\phi - \gamma )}
                                                               \cr
}
\eqno (17)
$$
This reduces to $\Phi_m$ when $\theta_{\bar m}  \rightarrow 0$ and
to $e^{i\gamma} \Phi_{\bar m}$  when $\theta_m \rightarrow \pi$ and,
in addition, we perform the rotation $\phi \rightarrow \phi + \gamma$.
So the Higgs field configuration in (17) will describe
a monopole and antimonopole connected by a $Z-$string
that is twisted through an angle $\gamma$ provided we choose the
gauge fields suitably. The gauge fields can be written down using the
general formalism developed by Nambu\refto{yn}:
$$
g W_\mu ^a = - \epsilon^{abc} n^b \partial_\mu n^c -
i cos^2 \theta_w
 n^a (\Phi^{\dag} ~ \partial_\mu \Phi - \partial_\mu \Phi^{\dag} ~\Phi )
\eqno (18a)
$$
$$
g' Y_\mu = - i sin^2 \theta_w
 (\Phi^{\dag} ~ \partial_\mu \Phi - \partial_\mu \Phi^{\dag} ~\Phi ) \ .
\eqno (18b)
$$
upto ``external'' electromagnetic potentials\refto{yn} and where,
$$
n^a \equiv  \Phi^{\dag} \tau^a \Phi \ ,
\eqno (19)
$$
is a unit vector.

The configuration in (17) and (18) describes a twisted segment of string
with twist angle $\gamma$. If we assume that $\gamma = 2\pi n/m$, where
$n$ and $m$ are integers,
we can join together $m$ of these twisted segments and form
a loop of $Z-$string that is twisted by an angle $2\pi n$. The Chern-Simons
numbers of this twisted
loop of string is easy to calculate using (7) and (8) - it is
$n N_F cos2\theta_w$. Now, dividing by the number of segments we had
joined together to form the loop, this yields
the Chern-Simons number of a segment twisted by an angle $\gamma$:
$$
CS = {{\gamma} \over {2 \pi}} N_F cos2\theta_w \ .
\eqno (20)
$$
This shows that we can find string configurations with arbitrary Chern-Simons
number by putting in a suitable amount of twist.
In particular, if we take $\gamma = \pi / cos2\theta_w$, the Chern-Simons
number is $N_F /2$ - the believed value for the
sphaleron\refto{minosbrihaye}. This leads us
to conjecture that the string with twist $\pi /cos2\theta_w$ will collapse
($\theta_m \rightarrow \theta_{\bar m} = \theta$) into the sphaleron for any
Weinberg angle.


We would like to point out that the above arguments are independent of the
existence and stability\refto{mjlptv} of the $Z-$string solution.
In our analysis we only needed {\it configurations} that look like linked
loops of $Z-$flux. However, if we were to consider the formation of
such linked fluxes during the electroweak phase transition, the existence
of $Z-$string solutions and the issue of their instability
would become important.

It has been suggested in the past that the sphaleron might secretly be a
configuration of electroweak strings\refto{nm, tv, mbtvmb, mhmj}. Our result
that linked loops of string can carry baryon number and the deformations
outlined above give substance to this belief. It may also be that other
knotted string configurations are equivalent to the sequence of solutions
conjectured in Ref. \cite{fujii} and Klinkhamer's $S^*$ solution\refto{fk}
could have an interpretation in terms of electroweak string knots with zero
linkage\refto{knot} since it is known that $S^*$ carries zero baryon number.

Next consider the possibility that electroweak strings were produced during
the electroweak phase transition. When such strings form, they will
be produced with some helicity. The question is: what is the helicity
density?

The answer to this question is likely to be very difficult and here we
will only attempt to answer a simpler question: what is the probability for
getting linked loops in a phase transition where $U(1)$ is broken completely
and topological strings are produced? It is simplest to think in terms
of a simulation with the algorithm used in Ref. \cite{tvav}. In this
algorithm, one throws down one of three $U(1)$ phases called $0,1,2$
(corresponding to the angles $0, ~ 2\pi /3, ~4\pi /3$) on the
vertices of a lattice. Now in traversing the links of the lattice,
the phase increases or decreases and this corresponds to traversing some
segment of the $U(1)$ circle. If on traversing the perimeter of a plaquette
and returning to the starting point, we traverse an angle $2\pi$ on the
$U(1)$ circle, we must have a string passing through the plaquette.
In this way, one can construct a whole network of strings on a lattice.
(For details see Ref. \cite{tvav}.)

This algorithm allows for the possibility of forming linked loops of string.
We have evaluated the probability for a small loop to be threaded by
another string within this algorithm and find it to be $\sim 10^{-4}$.
Therefore the helicity per unit volume is $\sim 10^{-4}/\xi^3$ where
$\xi$ is the correlation length at the time of string formation.

Finally, we wish to point out that the above
results suggest a scenario for the generation of baryon number in the
early universe\refto{ewbaryog}.
Suppose that a network of electroweak strings was produced
at the electroweak phase transition which then survived long enough
to fall out of thermal equilibrium. The network would consist of loops
and segments of electroweak string of which some would be linked and
twisted. The network will evolve and the helicity will change with time.
Every change in the helicity results in a baryon number change - somewhere
positive, somewhere negative. Now the evolution of the system
is governed by the full electroweak Lagrangian which is CP violating.
The CP violating terms would favour a change of helicity in one direction
over the other and hence baryon number would be produced. (Remember that
the change in the baryon number not only depends on the initial helicity
but also on the Chern-Simons number of the final vacuum.) If we use
the $U(1)$ string network results to estimate the number density of
helicity ($n_h$), we have $n_h (t_i) \sim 10^{-4}/\xi^3$.
If the (model dependent) CP violation bias parameter
that preferentially drives baryon number change in one direction is denoted
by $\epsilon$, the baryon number density produced will be:
$ \sim 10^{-4} \epsilon / \xi^3 $. For $\xi \sim T^{-1}$, the baryon
to photon ratio will be $\sim 10^{-4} \epsilon$. Granting all the
assumptions we have had to make, this estimate would agree with observations
only in particle physics models that give $\epsilon \sim 10^{-6}$.

\

\

\noindent {\it{Acknowledgements:}}

We are grateful to Manuel Barriola and Alex Vilenkin for discussions.
Some of this work was done while at the Aspen Center for Physics
and with support from the National Science Foundation.

\

\

\

\references

\refis{ewbaryog} Baryogenesis scenarios based on electroweak strings have
been proposed by
R. Brandenberger and A. C. Davis, Phys. Lett. B{\bf 308}, 79 (1993), and,
by M. Barriola, hep-ph/9403323 (1994).

\refis{comment2} We could also consider linked loops of
$W-$string\refto{tvmb, mbtvmb}. In that case, the analysis would be different
since the properties of $W-$strings are different from those of the
$Z-$string.

\refis{norm} $\Phi$ has been rescaled so that the vacuum manifold is
given by $\Phi^{\dag} \Phi = 1$.

\refis{ed} TV is grateful to Ed Copeland for help in clarifying this
point.



\refis{comment10} The last two terms in (6) give the linkage of the $Z-$flux
tubes with any electromagnetic fields that may be present. Note that the
term ${\vec A} \cdot {\vec B}_A$ is not present in (6).


\refis{tv1}  T. Vachaspati, Phys. Rev. Lett. 68, 1977 (1992);
69, 216(E) (1992); Nucl. Phys. B397, 648 (1993).

\refis{comment5} Perhaps the most graphic illustration of the immediate
reconnection of magnetic field lines in string intersections is to be found
in the case of ``peeling strings'' - see P. Laguna-Castillo and R. Matzner,
Phys. Rev. Lett. 62, 1948 (1989).

\refis{tv} T. Vachaspati, in proceedings of ``Texas/PASCOS 92:
Relativistic Astrophysics and Particle Cosmology''
(Ann. N. Y. Acad. Sci. Vol. 688, 1993).

\refis{mbgf} J. J. Moreau, C. R. Acad. Sci. Paris 252, 2810 (1961);
H. K. Moffat, J. Fluid Mech. 35, 117 (1969); M. Berger and G. Field,
J. Fluid Mech. 147, 133 (1984).

\refis{nt} N. Turok in ``Perspectives on Higgs Physics'', ed. G. Kane,
World Scientific, page 300 (1992).

\refis{yn} Y. Nambu, Nucl. Phys. B130, 505 (1977).

\refis{nm} N. S. Manton, Phys. Rev. D28, 2019 (1983).

\refis{fknm} F. R. Klinkhamer and N. S. Manton, Phys. Rev. D30, (1984)
2212.

\refis{review} A. G. Cohen, D. B. Kaplan and A. E. Nelson, to appear in
Annual Reviews of Nuclear and Particle Science, 43 (1993); UCSD-PTH-93-02;
BUHEP-93-4.

\refis{mhmj} M. Hindmarsh and M. James, ``The Origin of the Sphaleron
Dipole Moment'', DAMTP-93-18.

\refis{mjlptv} M. James, L. Perivolaropoulos and T. Vachaspati, Phys.
Rev. D46 (1992) R5232; Nucl. Phys. B395, 534 (1993).

\refis{tvav} T. Vachaspati and A. Vilenkin, Phys. Rev. D30, 2036 (1984).

\refis{tvmb} T. Vachaspati and M. Barriola, Phys. Rev. Lett. 69,
1867 (1992).

\refis{mbtvmb} M. Barriola, T. Vachaspati and M. Bucher,
``Embedded Defects'', Phys. Rev. D, to be published; TUTP-93-7.


\refis{knot} A well-known example of a knot with zero linkage is the
Whitehead link. See, for example, L. H. Kauffman, ``On Knots'', Princeton
University Press (1987).

\refis{tvintercom} T. Vachaspati, Phys. Rev. D 39, 1768 (1989).

\refis{fujii} K. Fujii, S. Otsuki and F. Toyoda, Prog. Theor.
Phy. 81, 462 (1989).

\refis{fk} F. R. Klinkhamer, Phys. Lett. B246, 131 (1990).

\refis{minosbrihaye} M. Axenides and A. Johansen, NBI-HE-93-74;
Y. Brihaye and J. Kunz, hep-ph/9403392 (1993).

\endreferences

\vfill
\eject

\head{Figure Captions}

1. A linked pair of loops and a delinking process (see text for details).
The parallel curves represent $Z-$magnetic field lines of the same string.



\endjnl
\end